\documentstyle[12pt,aaspp4]{article}
\newcommand{\gapproxeq}{
{\ \lower-1.2pt\vbox{\hbox{\rlap{$>$}\lower5pt\vbox{\hbox{$\sim$}}}}\ }}
\newcommand{\lapproxeq}{
{\ \lower-1.2pt\vbox{\hbox{\rlap{$<$}\lower5pt\vbox{\hbox{$\sim$}}}}\ }}
\begin{document}

\title{2MASS Studies of Differential Reddening Across Three Massive Globular Clusters}
\author{David R. Law\altaffilmark{1}, Steven R. Majewski\altaffilmark{1},
Michael F. Skrutskie\altaffilmark{1}, John M. Carpenter\altaffilmark{2},
and Hina F. Ayub\altaffilmark{1}}

\altaffiltext{1}{University of Virginia, Department of Astronomy, P.O. Box 3818,
Charlottesville, VA 22903: drlaw@alumni.virginia.edu, srm4n@virginia.edu,
skrutskie@virginia.edu, hfa9b@virginia.edu}

\altaffiltext{2}{Department of Astronomy, California Institute of Technology,
MS 105-24, Pasadena, CA 91125: jmc@astro.caltech.edu}

\begin{abstract}
$J$, $H$, and $K_S$ band data from the Two Micron All-Sky Survey (2MASS) are used to study the
effects of differential reddening across the three massive Galactic globular clusters $\omega$
Centauri, NGC 6388, and NGC 6441.  Evidence is found that variable extinction may produce false
detections of tidal tails around $\omega$ Centauri.  We also investigate what appears to be
relatively strong differential reddening towards NGC 6388 and NGC 6441, and find that differential
extinction may be exaggerating the need for a metallicity spread to explain the width of the red
giant branches for these two clusters.  Finally, we consider the implications of these results for
the connection between unusual, multipopulation globular clusters and the cores of dwarf spheroidal
galaxies (dSph).
\end{abstract}

\section{INTRODUCTION}
The giant Galactic globular cluster $\omega$ Centauri (NGC 5139, hereafter $\omega$ Cen) is not
only the most massive globular cluster in the Milky Way, but its stars have also been shown to have
a large spread in both age and metallicity (see for instance, many contributions in van Leeuwen et
al. 2002).  In light of these discoveries, it has been proposed (Lee et al. 1999, Majewski et al. 2000)
--- and now dynamically modeled (Tsujimoto \& Shigeyama 2003) ---
that $\omega$ Cen may be the nucleus of a disrupting dSph, similar to the Galactic globular cluster M54.
The second largest globular cluster, M54 is currently believed to be the core of the Sgr dSph
(Da Costa \& Armandroff 1995, Bassino \& Muzzio 1995, Sarajedini \& Layden 1995, Layden \& Sarajedini 2000)
and lies at a nucleation point for the several different metallicity
populations of Sgr (Sarajedini \& Layden 1995, Majewski et al. 2003).
The recent observation of apparent tidal tails around
$\omega$ Cen by Leon, Meylan \& Combes (2000, hereafter LMC00) bolsters the notion
of this cluster being the remnant of a  larger, now disrupted parent object, and, with $\omega$ Cen
 and M54 as models, Ree et al. (2002), Lee et al. (2001, 2002), and Yoon et al. (2000) 
have proposed that other clusters with
evidence of multiple populations, such as clusters with bimodal horizontal branches, may also be the
nuclei of disrupting dSph satellite galaxies.  Two clusters proposed to be M54 analogues
(Ree et al. 2002) are NGC 6388 and NGC 6441,
the third and fifth largest globular clusters in the Milky Way
respectively, since both have observed characteristics suggesting internal metallicity spreads.
Moreover, high mean metallicity clusters like these (see Table 1) should have only a
red stub of a horizontal branch
(HB), but NGC 6388 and NGC 6441 also have extended blue HB tails (Piotto et al. 1997, hereafter P97).
A {\it bimodal} (or extended) metallicity distribution may explain both this curious HB morphology (P97) as
well as the unusually wide red giant branches (RGB) observed in these clusters.  Because it is difficult to
understand how multiple generations of stars might be created in a stellar system of initial mass
of order $10^{5-6}M_\odot$ (which is not massive enough to retain significant amounts of gas
after an initial starburst), it is possible that the multi-population clusters are the remnants of 
{\it formerly} larger stellar systems, for example dSph galaxies, where multi-epoch
star formation histories are commonly found.  The verification of the existence of
additional multiple-population clusters would increase the need to formulate a general
process --- e.g., the disruption of initially larger systems --- that leads to a more substantial
\textit{class} of these objects.

These scientific questions are complicated by the presence of non-negligible differential reddening in
the fields of some the most interesting examples of potential multi-modal clusters.
In this paper, we concentrate on two effects of differential reddening relevant to the
globular cluster evolution picture outlined above: the potential creation of false signatures
of both tidal tails around, and of multiple metallicities within, globular clusters.
These issues are explored in case studies of the three massive globular
clusters $\omega$ Cen, NGC 6388 and NGC 6441.

Tidal tail searches around Galactic satellites can be particularly vulnerable to differential
reddening effects because search techniques typically rely on the detection of
star count overdensities.  Lines of sight with a lower dust
column density will naturally reveal more stars to a given magnitude limit.
This may create an
apparent overdensity of stars in regions with lower extinction that could be mistaken for, or
overemphasize, density enhancements from tidal tails.  In the case of the reported tidal tails around
$\omega$ Centauri by LMC00, the contours of the detected tails curiously and closely correspond to contours 
of the prominent IRAS $100\micron$ dust emission (and therefore, presumably, optical extinction by dust) 
in this field, as these authors have observed.  Because LMC00 did not correct for differential extinction 
in their optical starcounts (S. Leon, private communication), at minimum there
must be some dust-shaping of the morphology of
mapped $\omega$ Cen tidal tails.  We re-evaluate the possible impact of differential reddening
on large field studies of $\omega$ Cen by comparing a customized tidal tail search on raw 2MASS
data to one performed on 2MASS data differentially dereddened according to the dust maps of
Schlegel et al. (1998).

The cause for the observed width of the RGB in NGC 6388 and NGC 6441 has been variously
explained as differential reddening and/or the presence of multiple populations of different
metallicity, with no clear consensus emerging.  P97 originally concluded that differential
reddening on scales greater than eight arcseconds was unlikely for NGC 6388 but possible for NGC 6441,
although only to a few hundredths of a magnitude.  Later evidence indicating the presence of strong
differential reddening on small scales across both clusters was found by Raimondo et al. (2002)
using HST-WFPC2 small area surveys similar to those used by P97.  Pritzl et al. (2001, 2002) also
find differential reddening in NGC 6441 and NGC 6388 using CTIO $B$ and $V$ band photometry, although
they note that in NGC 6388 the differential reddening probably covers a smaller magnitude range.  If
differential reddening were significant, the reddening vector could reproduce the sloping red HBs
and spread the RGB stars perpendicular to the unreddened RGBs, inflating their apparent width.
Numerical experiments performed by Raimondo et al. (2002) indicate that differential reddening of
order $\Delta$E$(B-V) \approx$ 0.1 would be substantial enough to cause an unreddened HB of similar
metallicity (e.g., that of 47 Tucanae) to resemble a sloping red stub similar to that of NGC 6388 or
NGC 6441.  Nevertheless, Raimondo et al. conclude that reddening is not the reason for the strange HB's
since their sloping nature is present even in small area subsamples exhibiting a thin RGB, which
suggests that a second effect is at work.

Pritzl et al. (2001, 2002) and Ree et al. (2002) argue that variable metallicity may
adequately explain the observed features of the NGC 6388 and NGC 6441 CMDs,
and recent stellar evolutionary models (e.g.,
Sweigart \& Catelan 1998) confirm that a metallicity spread can naturally create a sloping HB.
Specifically, Ree et al. (2002)  find that two distinct populations separated by 1.2 Gyr and 0.15 dex in [Fe/H]
can reproduce both the observed wide RGBs and tilted HBs in the clusters.
However, Pritzl et al. (2001, 2002) argue that the
simple two-population model fails to reproduce the observed HB slope and that a more general
metallicity \textit{spread} is necessary.

Because selective extinction at the 2MASS wavelengths $J$ (1.25 $\micron$), $H$ (1.65 $\micron$),
\& $K_s$ (2.17 $\micron$, hereafter simply referred to as $K$) is roughly one tenth that in
visual bands, the magnitude of differential reddening at these wavelengths is commensurately reduced.
Thus, we apply the 2MASS point source catalog to test whether: (1)
differential reddening affects interpretations of an {\it infrared} search for tidal tails around $\omega$
Cen,
and (2) the NGC 6388 and NGC 6441 RGBs are broadened at near infrared wavelengths
where differential reddening will have less impact on the width of the cluster RGB's.

\section{DATA}
$J$, $H$, and $K_S$ band data were drawn from the 2MASS all-sky point-source working survey
database as of July, 2002.  Initial
values of the mean field extinction,  $E(B-V)$, and distance modulus, $(m-M)$, for each cluster were
obtained from the Harris compilation as posted on the World Wide Web
(http://physun.physics.mcmaster.ca/Globular.html) on June 22, 1999 (see Harris 1996).
These were converted to the required infrared extinction parameters using the standard
extinction models of Cardelli, Clayton, and Mathis (1989).  Table 2 gives a complete
list of the adopted extinction parameters.

\section{$\omega$ CENTAURI: A 2MASS TALE}

\subsection{Tidal tail search method}
The LMC00 study of optical starcounts in the $\omega$ Cen field makes
no corrections for differential reddening because
magnitude calibrations were not available for the majority of clusters in their study
(S. Leon, private communication).
As Figures 1a and 1b illustrate, the proposed tidal tails
of $\omega$ Cen are from enhanced star counts whose pattern appears
to follow regions of lower dust
absorption on the IRAS $100\micron$ map.  This effect is particularly noticeable on the western
edge of the cluster.  LMC00 note and discuss the possibility that the star counts are affected
by reddening, but determine that the tidal tails are of a sufficiently large S/N that
differential reddening should not significantly affect their tidal tail search.

By comparing starcount patterns obtained using raw 2MASS data with
those from 2MASS data differentially dereddened according to the dust maps of Schlegel et al. (1998) we intend
to re-evaluate whether uncorrected differential extinction can
produce a false tidal tail signal.
The 2MASS data has a $99\%$ completeness limit
at $J = 16$ mag in the direction of $\omega$ Cen, yielding
almost 500,000 stars within a $6^{\circ} \times 6^{\circ}$ box centered on the cluster.
Because a photometric search for diffuse tidal tails presents considerable challenges, not least of which is
substantial field star contamination in this low latitude field, a two stage method
is employed to minimize field star contamination and 
probe for starcount excesses as deeply into the surrounding field as possible:  First,
a coarse filtering mask in color-color-magnitude (CCM) space is used to extract only those stars
whose CCM location is similar to that of stars observed at radii less than the cluster core
radius.  Although this coarse mask does not take into account foreground/background
contamination of the
cluster core, it filters out a significant number of stars whose CCM locations are significantly
different from those in the cluster.
Second, we then apply a signal-to-noise filter similar to that used by Odenkirchen et al. (2001)
to separate cluster stars from field stars and emphasize starcount overdensities in the potential tidal tails.

In the first stage of filtering, a filtering mask is applied in CCM space to pre-select
those CCM locations containing stars within 100 arcseconds of the center of the cluster
(approximately $2/3$ the core radius $r_c$, see Table 1).  We divide this irregularly shaped
region in ($J$, ($J-H$), ($J-K$)) space into a $50\times 50\times 50$ grid in CCM space bounded by the
limits of the CCM region; this grid size is selected to achieve a balance between coarse binning
and low number statistics.  With the individual cells described by indices ($i$, $j$, $k$)
respectively, we define
$n_C(i,j,k)$ as the number of stars within 100 arcseconds of the cluster center that fall within
each grid cell in CCM space.  To obtain an estimate of the number of
field stars included in $n_C(i,j,k)$, we
represent the field star distribution, $n_F(i,j,k)$, by the stars within an annulus
$100\farcm0 \leq r \leq 150\farcm0$ (roughly three times the tidal radius of the cluster).
We assume that the distribution of
field stars is relatively even across the cluster and scale the field mask by the ratio $q$ of
the areas over which the field and core regions were summed.  The selection mask $P$ is defined
as the ratio of the core mask to the sum of the core and scaled field masks.
\begin{equation}
P(i,j,k) = \frac{n_C(i,j,k)}{n_C(i,j,k)+q^{-1}n_F(i,j,k)}
\end{equation}
$P(i,j,k)$ will have a value near one for regions of CCM space where
the core sample $n_C(i,j,k)$ is not substantially contaminated with field stars, and will decrease
as contamination becomes more severe.  The complete data set is filtered using this CCM
selection mask:  A star is rejected from the data set if the corresponding $P(i,j,k)$ is less
than some minimum value $P_{min}$.   This algorithm is a slightly more sophisticated means
to eliminate stars unrelated to $\omega$ Cen than simple preliminary cuts in the color-magnitude
(CM) plane.
We find that a value of $P_{min} = 0.9$ allows us to select cluster stars over field stars
most optimally and appears to preserve the CMD of the cluster center best.  This preliminary
filtering rejects roughly 90\% of the stars in the original data set.

In the second stage of filtering, we apply a modified version of Grillmair et al.'s (1995) method
in the CM plane as described by Odenkirchen et al. (2001).  This second stage
eliminates field stars more effectively than the first stage, but we find that the method does not
perform well unless a coarser pre-selection of stars is performed first to reduce the data set to a
size that the more
computationally complex second-stage routine can handle.  With some recycling of notation,
we define a grid in $(J, (J-K))$ space, with indices ($j$, $k$) respectively, and find the signal-
to-noise ratio ($S/N$) of probable cluster stars to field stars in each CM grid cell.  We count the number
of stars $n_C(j,k)$ that fall within each grid cell and lie within $r=20\farcm0$ of the cluster
center (chosen to correspond to the region in which LMC00 observe circular cluster surface density 
contours, see Fig. 1a), and the number of stars $n_F(j,k)$ that fall
within the CM grid cell and lie between $r=100\farcm0$ and $r=150\farcm0$ arcminutes from the
cluster center.  The field star population is assumed to be uniform, and the expected number of field stars of a
given ($j$, $k$) that fall within $r=20\farcm0$ of the cluster core is subtracted from $n_C(j,k)$.
With $q$ representing the ratio of the area of sky on
which $n_F$ and $n_C$ respectively have been defined, the local S/N ratio of cluster to field stars for
each grid cell ($j$, $k$) is, by standard Poissonian statistics,

\begin{equation}
s(j,k)=\frac{n_C(j,k)-q^{-1}n_F(j,k)}{\sqrt{n_C(j,k)+q^{-2}n_F(j,k)}}
\end{equation}

It is possible to use $s(j,k)$ as a mask for selection in the CM plane by eliminating all $(j,k)$
grid cells with $s(j,k)$ less than some limiting value $s_{lim}$, which is chosen such that the
number of stars in the tidal tail region (as reported by LMC00) is a maximum relative to the
field star population.
The new matrix $n_T(j,k)$ represents the total number of stars in each $(j,k)$ grid cell for the
region\footnote{We describe this
region as the two sections of an annulus (of inner radius $50\farcm0$
and outer radius $100\farcm0$) located directly north and south of the circle
described by the inner radius.} of the cluster in which tidal tails have been reported by LMC00 (see Fig. 1a), 
as shown by the dashed lines in Fig. 1a.
Starting from $s_{lim}=s_{max}$ (where $s_{max}$ is the maximum value of all $s(j,k)$)
and iterating downward to $s_{lim}=0$, we calculate $N_{tail}=\sum_{j,k} n_T(j,k)$
and $N_{field}=\sum_{j,k} n_F(j,k)$ for all ($j$, $k$) such that $s(j,k) \geq s_{lim}$.
Hence, $N_{field}$ and $N_{tail}$ respectively are the total number of field stars and cluster stars in LMC00's tidal tail
regions whose locations in $(J,(J-K))$ space satisfy $s(j,k) \geq s_{lim}$.  The optimal value of $s_{lim}$
is that which maximizes $N_{tail}/N_{field}$; hence this final selection mask filters the data set for
all stars which fall within cells in CM space such that $s(J,(J-K)) \geq s_{lim}$.

The positions of stars from this filtered set
are then converted to a smooth density distribution by application of a two-dimensional Gaussian
kernel with an adaptive half maximum width (e.g., Silverman 1986) for each star set to the distance
of the 30$^{th}$ nearest neighbor in the plane of the sky.  Each of these Gaussians is
evaluated at finely spaced grid points in the plane of the sky, and these values are summed to
create an image of relative stellar densities at each location on the grid (Fig. 1c and 1d)
\footnote{The reduced field of view of these figures is chosen to correspond to that of LMC00.}.

We find that this $S/N$ optimization method worked efficiently to reduce the 2MASS data set
by a further 80-90\% and brought out starcount overdensities around the cluster that have the
appearance of tidal tails.  Focusing the definition of $n_T$ on the regions found by LMC00 to
possess tidal tails is considerably more effective at bringing these apparent tails out of the
background than defining $n_T$ more generally; however, this particular approach
obviously is not useful
for a general survey for previously unknown tidal tail searches because it requires
\textit{a priori} knowledge of where tails may be found.  We verified that the overdensities
found by this method are not merely artifacts of the definition of $n_T$ in the region of the
LMC00 expected tails by redefining $n_T$ in locations rotated around the cluster core by 90 degrees
(that is, protruding east-west as opposed to north-south).  Using this definition of $n_T$,
no starcount overdensities were detected around the cluster, which supports that the apparent
tails found with the first $n_T$ definition had not been an artificial product of the
processing method and that the method of sampling $n_T$ in the tail region is a crucial
step in bringing those tails out of the background noise.

\subsection{Results}
In the analysis of the raw 2MASS data described above a starcount excess to the north and
south of the cluster is found.  As Fig. 1c shows, to within the limits imposed by the
smoothing kernel, the location of the 2MASS tails corresponds
to that of the tails found
by LMC00.
%

The same analysis has been repeated with 2MASS data dereddened according to the extinction maps of
Schlegel et al. (1998)\footnote{We have used the programs available on the World Wide Web at
http://www.astro.princeton.edu/\~{}schlegel/dust/index.html}.  The Schlegel et al.
maps have a resolution of
approximately 6.1 arcminutes, which corresponds to about 1500 cells of reddening data over a
$4^{\circ} \times 4^{\circ}$ field of view,
and show differential reddening of order
$\Delta E(B-V)=0.18$ mag over that field of view.
In contrast to what is found for the raw 2MASS data, the results of the analysis on the
dereddened 2MASS data do not reveal any elongation of the circular cluster contour lines or
any overdensities resembling tidal tails (Fig. 1d).  Fig. 2 plots radial surface density
profiles for both original and dereddened 2MASS data, and shows the surface density of the
dereddened cluster to fall off smoothly with radius, like a King (1966) profile.  In contrast,
the surface density of stars in the non-dereddened cluster data appears to drop at $\log(r)=1.8 $
($r \approx$ 60 arcmin) --- corresponding to the low starcount regions at this radius to the east
and west of the cluster --- and rise again at $\log(r)=2.0$ ($r = 100$ arcmin) --- which corresponds to
high starcount regions to the north and south of the cluster (Fig. 1c).

Given the fidelity with which the apparent infrared and optical tails follow the regions of
lower reddening on the IRAS 100 micron map (Fig. 1b, particularly on the western and southern
edges of the cluster), and the disappearance of the infrared tails in the dereddened cluster
data, we speculate that by analogy the $\omega$ Cen tidal tails previously detected with the
optical starcounts, which are more affected by dust extinction, could have been artifacts of
differential reddening.  We must also point out, however, that our test of the 2MASS database is not
completely analagous to the LMC00 analysis:
Due to the $J = 16$ mag completeness limit of 2MASS
we are unable to probe to the same effective depth as the LMC00 study.  The differences
in relative depth are demonstrated by the fact that while LMC00 observe reasonably
circular contours out to a radius of about 45 arcminutes (roughly the tidal radius, see Table 1),
we only find circular contours out to about 40 arcminutes before we begin to observe distinct
elongation of the contours into potential tidal tails (Fig. 1).  Thus, the Leon et al. counts
reach fainter surface brightness limits than does 2MASS and are sensitive to more
diffuse, lower density tails than are we.  Nevertheless, we regard the 2MASS test presented here
as a cautionary example demonstrating that a proper accounting of differential reddening is an 
important step in mapping any tidal tails in the $\omega$ Cen (or any other heavily obscured) field.

\section{DIFFERENTIAL REDDENING IN THE FIELDS OF NGC 6388 AND NGC 6441}
\subsection{Processing}

To investigate the effects of differential reddening on the width of the RGB in the clusters
NGC 6388 and NGC 6441 we use the uniform photometry of 2MASS to look for spatial patterns in the
distribution of red giant star colors within each cluster.  47 Tucanae (NGC 104, hereafter 47 Tuc)
is adopted as a control case for a cluster of similar mean metallicity but with little differential
reddening (average $E(B-V) \approx$ 0.02).  We also compare our results to the $\omega$ Cen data ---
although this cluster has an overall [Fe/H] much lower than the other clusters, it nonetheless
provides a useful example of the effects of \textit{variable metallicity} within a cluster over
spatial scales with relatively little differential reddening on the scale of arcminutes
($\Delta E(B-V)=0.01$ mag, note that the previous section explored extinction variations in
$\omega$ Cen on the scale of {\it degrees}).

\subsubsection{($J-K$) Color Distributions}
Analysis of RGB color \textit{spreads} at different magnitudes in one cluster as well
as differences between clusters is simplified if we remove
the mean RGB color-magnitude relation for each cluster (i.e. make the center of the RGB locus for each cluster appear
vertical in the color-magnitude diagram).
Since the [Fe/H] of the unreddened cluster 47 Tuc is
similar to that of NGC 6388 and NGC 6441 to within about 0.25 dex (see Table 1), the shape, height and
position of the RGBs for the three clusters in a reddening corrected, absolute color-magnitude
space should be similar.  Hence, we use the well-defined 2MASS CMD of 47 Tuc to ``straighten''
the RGB of the clusters in our sample (Figure 3a).  We convert apparent to absolute magnitudes
using the distance moduli in Table 2 and consider only that part of the RGB with absolute
magnitude $M_K \leq -2$ to minimize contamination from cluster horizontal branch (HB) as
well as field stars.  A Legendre polynomial is fit to the 47 Tuc RGB, with a rejection
threshhold of 2$\sigma$ iterated until convergence.  The final polynomial is converted to
power series coefficients up to third order, which are used to straighten the RGBs of all four
clusters.  We define an adjusted $(J-K)$ color given by:
\begin{equation}
(J-K)_A = (J-K)-E(J-K)-(a+b\times M_K+c\times M_K^2)
\end{equation}
where $E(J-K)$ values are given in Table 2, and the coefficients $a$, $b$, and $c$ are 
power series coefficients
given in Table 3.  Similar equations are derived for the $((J-H),K)$ and $((H-K),K)$ CMDs.

Using the adjusted values, histograms of colors readily show that the RGBs of NGC 6388
and NGC 6441 (Figure 4) are wider than those of 47 Tuc and even that of $\omega$ Cen, which is well
established to have a metal spread from [Fe/H] $\approx$ -2.2 to -0.7  (Vanture, Wallerstein,
\& Suntzeff 2002).  The latter result is somewhat surprising, given the fact that the
metal-rich RGB of 47 Tuc provides a poor approximation to the metal-poor RGB of $\omega$ Cen
and produces a clear tilt in the ``adjusted'' $\omega$ Cen RGB (Fig. 4).  The Fig. 4 histograms are
divided into five $(J-K)_A$ colors bins in order to study the spatial distribution of RGB stars by color.
The five color bins from the lowest value of $(J-K)_A$ (most blue) to
the highest (most red) are referred to as bins A through E, respectively.
Because globular clusters are
dynamically mixed on timescales that are short compared to chemical enrichment,
there should be no significant difference in the spatial distribution of the
stars in these five bins.  For example, we show below (Section 4.2.1) that
neither 47 Tuc nor $\omega$ Cen
exhibit any such spatial variation.  On the other hand, if differential extinction were significant,
the distributions of stars by color would be uneven across the face of the cluster as a reflection
of reddening variations.  This is a test we now apply to NGC 6388 and NGC 6441.

Various methods to divide the sample population into color bins were explored.  The results are
not very sensitive to which method is chosen.  The following technique ensures similar treatment
of each cluster: Bin ranges are assigned to each cluster using a $(J-K)_A$ histogram of a subsample
of the purest cluster stars, defined by an angular radius where the density of stars decreased to
a limiting density given by:
\begin{equation}
Limiting\:Density = (1/e^3)\times(Central\:Density - Outer\:Density)
\end{equation}
(where the outer density is determined as the density of stars at the edge of the field of view).
In all cases, this resulted in a sample of stars within a radius of between roughly one and three
arcminutes, which is outside the core radius but well within the tidal radius of each cluster (see
Table 1).  Using a color histogram with binning width 0.02 mag, the central three bins are
defined within the $(J-K)_A$ range from the central peak to the colors where counts decrease to
$1/e$ of the central peak.  Bin C is centered at the peak of the histogram and has 1/3 of the
width of B+C+D.  The range of bin A is determined as the color range blue-ward of bin B up to
where the counts first drop to one (to eliminate extreme colored stars from the sample).
Likewise, bin E is defined as the range red-ward of bin D up to where the counts first drop to
one (Fig. 5).  Field star contamination (or contamination by non-RGB stars in the cluster),
not differential reddening, is assumed to be the reason for stars that
lie very far from the fiducial RGB, and the described culling of the data set promises to retain only those
stars that make up the bulk of the RGB and that could potentially have been scattered to their
current locations in the CMD by differential reddening.  These color bin definitions are
applied to the full 300 arcsecond data set.  Similar binnings were created for $(J-H)_A$ and
$(H-K)_A$ histograms, although
only the $(J-K)_A$ and $(J-H)_A$ distributions were found to be useful in our analysis
(see below).

\subsubsection{($J-H$) and ($H-K$) Colors}

Before proceeding further, we apply one additional filter designed to preserve reddening
effects and reduce noise from
astrophysical and photometric scatter.  This filter takes advantage of the fact that the degree of
scatter due to reddening in one color should be correlated to that in another color.   A
Legendre polynomial with an iterative $2\sigma$ rejection threshhold (similar to that used
earlier in Section 4.1.1) was fit to the cluster $(J-K)_A$ vs. $(J-H)_A$ color-color diagram
(Fig. 6), and color-color cuts were performed to eliminate stars at greater than $2\sigma$ from the
Legendre fit.  This independently derived fit for each cluster is very close to the fiducial reddening
vector for standard extinction models with $R_V = 3.1$, lending further evidence to support the
presence of differential reddening across each cluster.  Figs. 6 and 7  show
color-color plots of $(J-K)_A$ versus $(J-H)_A$ and $(H-K)_A$ color, respectively, with the
adopted outlier cuts shown.

Note that the dynamic range in color of $(H-K)_A$ is small compared to the reddening effect
sought, and it is difficult to determine which stars appear redder than average as a result
of differential reddening instead of photometric errors or astrophysical reasons.  We
therefore reject $(H-K)_A$ color from our analysis, since it does not have the discriminating
power required to pick out the required distribution from the noise of points in Figure 7.

\subsection{Results and Discussion}
\subsubsection{Five Bin Results}

Spatial plots of the stars in each of the $(J-K)_A$ bins for 47 Tuc, $\omega$ Cen, NGC 6388 and
NGC 6441 are shown in Figures 8 - 11 respectively.  As expected for a monometallic population
with little reddening, 47 Tuc shows no spatial segregation of the stars in the five bins.
Although giant stars in $\omega$ Cen have been found to exhibit a spatial variation in
metallicity (Jurcsik 1998), this variation is on scales of roughly 20 arcminutes; in contrast, the
mixed metallicity population in our smaller field of view shows no spatial segregation by RGB
color.  Both 47 Tuc and $\omega$ Cen have few stars in bins A and E relative to their total population.
In contrast, NGC 6388 exhibits a pronounced abundance of redder stars (bins D and E) in the northern
and northwestern regions of the cluster compared to southern regions (Fig. 10), and the extreme
bins A and E each contain significantly more stars relative to the 
total population of the cluster than these bins contained for 47 Tuc and $\omega$ Cen.  Comparison
to the 47 Tuc and $\omega$ Cen templates for monometallic and multi-metallic clusters (respectively)
having comparatively little differential reddening over small scales leads one to the conclusion that
it is unlikely that the
NGC 6388 color spread is due primarily to metallicity effects.  Rather, Fig. 10 suggests that
differential reddening on small scales is causing (1) a greater number of stars to appear in
bins A and E, and (2) large spatial variations by observed stellar color.

Likewise, NGC 6441 (Fig. 11) shows an enhancement of redder stars (bins D and E) in the southeastern
quadrant of the field of view, supporting the view that it too is subject to substantial differential
reddening.  Using isochrone fitting in subfields, Heitsch \& Richtler (1999) similarly find
that the southeast region of NGC 6441 has higher $E(B-V)$ than the northwestern region by about
0.15 mag.  Layden et al. (1999) concur with Heitsch \& Richtler to within a few hundredths of
a magnitude using RR Lyrae stars and CMDs in sectors of an annulus around the cluster center.
However, in contrast to the distinct dichotomy in NGC 6388 (redder stars in the northwest, bluer
stars in the southeast), the spatial distribution of stars in bins A through C in NGC 6441 appears
relatively constant (with the possible exception of decreased counts in the southeastern
quadrant of bin A), and the cluster presents the appearance of an inhomogeneously distributed
redder population superposed on a centrally concentrated bluer population background.
This effect may be an artifact of the severe contamination of the cluster by field stars
since NGC 6441 is located in an extremely crowded field near the Galactic center (as vividly
demonstrated by the washed out starcount map shown in Figure 12).  Indeed, Heitsch \& Richtler (1999)
note the strong contamination of their own CMD of NGC 6441 by field stars that cover the lower
part of the cluster RGB.  In this case, strong contamination by blue field stars may contribute
to the apparent width of the cluster red giant branch.

\subsubsection{Reddening Maps}
Figure 12 shows 2MASS starcount density (Panel a)
and mean stellar color (Panel b) maps for all four of our clusters.  In this figure, higher starcounts
and redder mean colors appear whiter (Table 5 provides the numerical values for
each greyscale).  The field of view of these maps is more
than 100 times larger than that used for the analyses in the rest of this section since this method only
provides coarse resolution maps which, however, are useful for
understanding qualitatively the global character and degree of
patchiness of the dust obscuration around each cluster.
For example, assuming that the average intrinsic star
color in a given direction is relatively uniform over degree scales, the mean color maps can be
interpreted as showing the general character of the reddening in the direction of each cluster.
However, these maps have too coarse a greyscale to show color variations of order $E(B-V) \leq 0.1$.

Fig. 12 shows that the mean stellar color in the direction of NGC 6388 is not only patchier than
that in the direction of 47 Tuc or $\omega$ Cen but that the number of redder stars appears to
be greater to the north of the cluster than to the south
\footnote{Note that this remark applies to the central few pixels immediately around the cluster.}, 
in agreement with the trend found
above.  Distinct regions of higher reddening and lower reddening are demarcated by a boundary
that passes close to the center of the cluster.  Differential reddening across this apparent
dust boundary may produce the observed overabundance of redder RGB stars in the north -
northwestern regions of the cluster compared to the south - southeastern regions.

The NGC 6441 field mean color map also shows evidence of patchy extinction, although it is not
obvious whether any of these patches could produce the observed reddening patterns near the
cluster that are likely the case for NGC 6388.  However, the presence of large scale patchiness
is sufficient to suggest the possibility of patchiness on the angular scales explored in Section 4.2.1.

\subsubsection{Kernel-smoothed color distributions}
Histograms of star colors within regions of greater reddening should be offset from those of
subregions with lesser reddening.  To study the color distributions in subsections of
the cluster fields, which will have larger statistical noise,
we apply a uniform width-normalized Gaussian kernel (e.g., Silverman 1986)
to the $(J-K)_A$ color distribution of stars in NGC 6388 and NGC 6441.  Figure 13 shows these smoothed
distributions for the northwest and southeast quadrants of each cluster.  The half maximum width
of the kernel is the minimum value that produces a clearly defined central peak in the distribution
for a given cluster.  NGC 6388 required this smoothing width to be only 0.02 mag, while NGC 6441
required a wider Gaussian of width 0.06 mag to smooth out secondary peaks adequately.

The central peak of the distribution of the northwestern quadrant of NGC 6388 is offset from that
of the southeastern quadrant by roughly $\Delta(J-K)_A =$ 0.061 mag, and the peaks of the
northwestern and southeastern quadrants of NGC 6441 are offset by roughly $\Delta(J-K) =$ 0.072 mag.
Because the width of the kernel used for NGC 6441 is of order the offset between the peaks, the
precision of the derived $\Delta(J-K)_A$ is less for NGC 6441 than for NGC 6388.  Nevertheless, the
$\Delta(J-K)_A$ results for both clusters are in agreement with the qualitative arguments for
differential reddening presented earlier in Sections 4.2.1 and 4.2.2, and our value for the
differential reddening across NGC 6441 is consistent with the more detailed optical extinction maps
produced by Layden et al. (1999) and Heitsch \& Richtler (1999) to within 0.02 mag in $E(J-K)$.
If such an offset exists between the color distribution of stars in different regions of each
cluster, then the artificial width produced by the superposition of these offset distributions
may contribute to the thicker RGB of the entire cluster.  Fig. 14 simulates the effect of
superimposing two offset color distributions by artificially reddening the northern half of the
monometallic cluster 47 Tucanae by 0.061 magnitudes in $(J-K)_A$; it is observed that while
such a superposition can artificially inflate the width of the color distribution (Fig. 13,
lower panels), it is only by about half the amount required to re-create the width of the
distributions observed in NGC 6388 and NGC 6441.
The inability of this simulation to entirely explain
the RGB widths of these clusters may be due to the presence of a large reddening \textit{spread}
(as opposed to the simplistic two value model used here), or it may be indicative of a contributing
effect of differential metallicity.

\subsubsection{Kolmogorov-Smirnov Statistics}
To test the significance of the differences in $(J-K)_A$ color distribution in different quadrants
of each cluster, we applied the Kolmogorov-Smirnov (KS) test to compare spatial color distributions
numerically.  As a calibration of the method, we first compared the color distribution across
the clusters 47 Tuc and $\omega$ Cen.  As Tables 6 and 7 show, the $(J-K)_A$ color distribution
of stars in, for example, the northwestern quadrant of the clusters correlates with the
distribution of stars in the southeastern quadrant with significance levels of about 0.5 and 0.25
for 47 Tuc and $\omega$ Cen respectively.  These significance levels are consistent with a
probability that stars within the two regions could have been randomly drawn from one parent
distribution.  Tables 6 and 7 demonstrate that this relation significance is roughly similar for
comparisons between all quadrants for both 47 Tuc and $\omega$ Cen.

Correlations between the southeastern and southwestern, and northeastern and northwestern
quadrants of NGC 6388 have significance levels of roughly 0.9 and 0.5 respectively.  In contrast,
the KS test yields a probability of only 0.002 that the northwestern and southeastern quadrants
share a similar color distribution (Table 8).  Compare this to the variable metallicity
cluster $\omega$ Cen, which maintains a relatively high correlation significance of color
distribution between all quadrants in spite of its metallicity spread.  It appears unlikely that
a metallicity spread is the dominant cause of the RGB width in NGC 6388 (we expect mixing in
clusters in any case).  Instead, the stars in these regions almost certainly come from
statistically distinct color populations, pointing again to the presence of strong differential reddening.

Application of the KS test to the $(J-K)_A$ color distribution in the southeastern and
northwestern quadrants of NGC 6441 yields a near zero likelihood of correlation (Table 9).
Qualitatively, this is not surprising because there are very few stars from bins D and E in the
northwestern quadrant while there are a considerably higher number of redder stars that lie in
the southeastern quadrant.  However, while these two quadrants are by orders of magnitude the
least correlated regions of this globular cluster, correlations between all other combinations
of quadrants are only of order 0.01 to 0.1, which suggests (as we have concluded in Section 4.2.2)
that reddening in NGC 6441 is patchy on smaller angular scales
than in NGC 6388.

If differential reddening is producing the observed width in these clusters, the RGB histograms
in sub-regions of each cluster should also be narrower than that of each cluster as a whole.
Using HST-WFPC2 visual band data for NGC 6388 and NGC 6441, Raimondo et al. (2002) find this to be
the case for both clusters.  In summary, it appears very probable that differential reddening
is artificially inflating the apparent thickness of the RGB in these two clusters, although
the thickness may not be fully explained by this effect alone.

\section{DISCUSSION AND CONCLUSIONS}
Differential extinction complicates photometric studies of Galactic stellar populations, and
even nearby globular clusters with $E(B-V)$ as low as about 0.1 mag may be visibly affected.
In our search for tidal tails around the giant globular cluster $\omega$ Cen we have found that
differentially dereddening 2MASS photometry largely eliminates apparent tidal tail candidates
present in data that have not been corrected for extinction.   Given the similarity of the false
2MASS tidal tail signature to the optical tails reported by Leon et al, it would be beneficial to
verify the diffuse optical starcount distribution in the $\omega$ Cen field with deep visual
photometry that accounts for the $\Delta E(B-V)$ variations.

We have also investigated the effects of differential reddening on the CMDs of the clusters NGC
6388 and NGC 6441, which exhibit abnormally large spreads in their RGBs.  The analyses presented
make use of spatial maps by stars in color bins, maps of mean stellar color, kernel-smoothed
histograms of stellar color in sub-regions, and KS statistical analyses of the distribution of
($J-K$) and ($J-H$) colors across the clusters.  These tests indicate that the northwestern
quadrant of NGC 6388 is subject to about $\Delta E(B-V) \approx$ 0.12 mag more reddening than
the southeastern quadrant, and that the southeastern quadrant of NGC 6441 is more heavily reddened
by about $\Delta E(B-V) \approx$ 0.14 mag than the northwestern quadrant.  These findings are
consistent with previous results from Heitsch \& Richtler (1999) and Layden et al. (1999).  As
numerical experiments performed here and by Raimondo et al. (2002) have shown, this amount of
differential reddening could produce anomalously wider red giant branches for
these clusters (in addition to the tilt observed in their horizontal branches, as demonstrated
by Raimondo et al.).  Raimondo et al. have shown that the RGB in sub-regions of
these clusters is thin (as expected if differential reddening were the cause of the width of
the RGB across the entire cluster).  Athough we find that possible width inflation due to the superposition of thin
RGBs from two {\it quadrants} subject to different amounts of differential extinction may not be sufficient to
completely explain the observed width of the cluster RGBs, it may be possible that the RGB width could be explained
more fully by a more realistic superposition of RGBs from smaller subregions.
Therefore, if not accounting for the entire RGB-widening effect, at minimum differential reddening may be
exaggerating metallicity-induced RGB spreads and the actual spread in age or metalliticy
within the clusters is likely not as large as has previously been suggested
(e.g., Pritzl et al. 2001, 2002).

The number of known globular clusters showing clearly anomalous metallicity spreads remains small.
With few bona fide members, this ``class'' of globular cluster can still comfortably be accomodated
by origin pictures that are exceptional, e.g., along the lines of the M54 - Sgr paradigm.  However,
even in the obvious case of the multi-population $\omega$ Cen, the M54 analogy may be less
compelling than previously thought, and the picture that some globular clusters are the ``residue''
of dSph disruption more ambiguous.

\acknowledgments
The authors would like to thank S. Leon and G. Meylan for helpful discussion regarding the
effects of reddening on $\omega$ Centauri.  This publication makes use of data products from
the Two Micron All Sky Survey, which is a joint project of the University of Massachusetts and
the Infrared Processing and Analysis Center, funded by the National Aeronautics and Space
Administration and the National Science Foundation.

\clearpage
\begin{figure}
\plotone{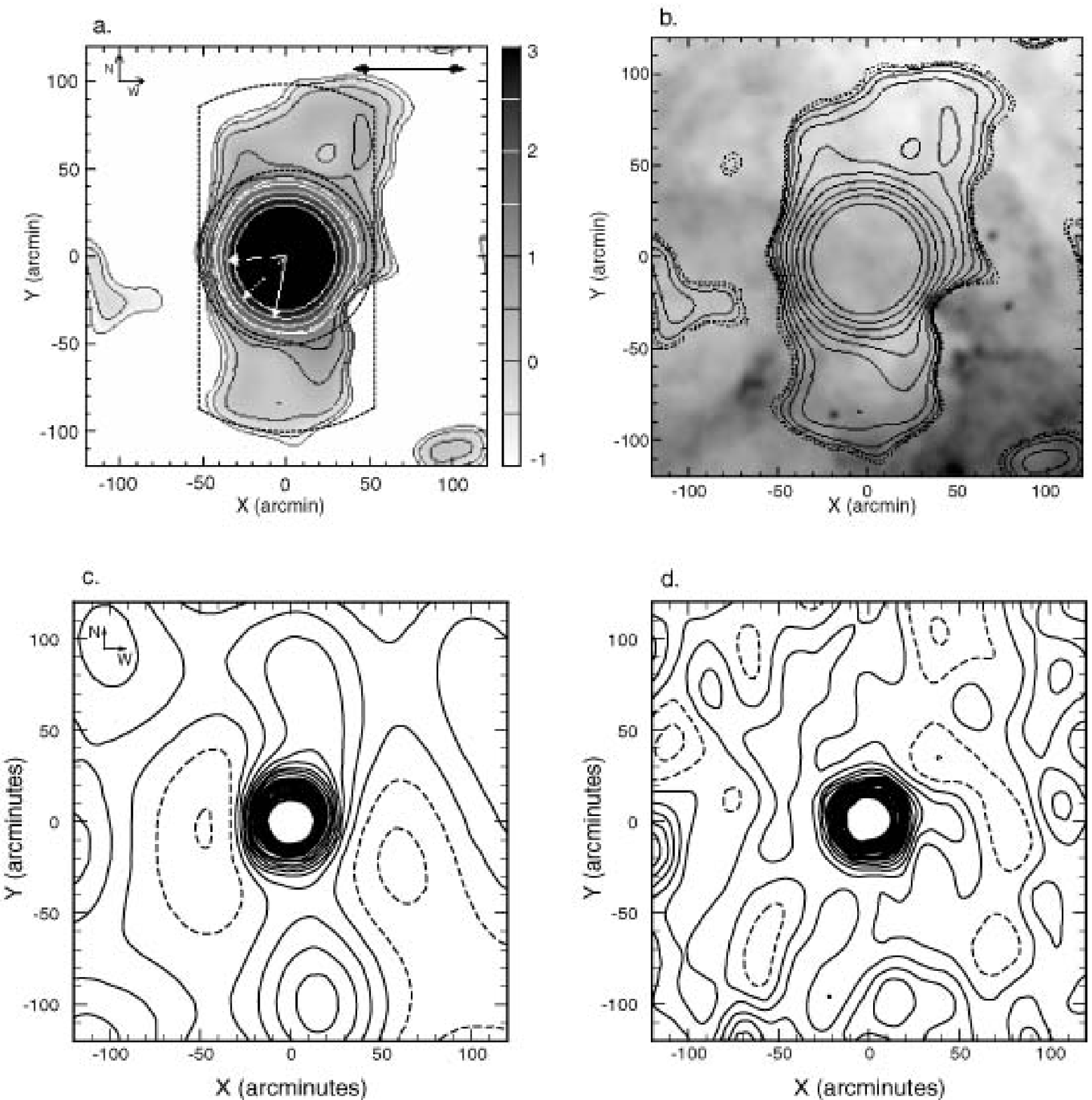}
\caption{(a): Smoothed function surface density plot of $\omega$ Centauri showing tidal tails (using 
logarithmic contours)
found by Leon et al. (2000).  The dotted line describes the region within which $n_T$ is defined
(Section 3.1), the dotted arrow represents the cluster proper motion, the dashed arrow the direction
of the Galactic center, and the solid arrow the direction perpendicular to the Galactic plane.
The horizontal double arrow represents a length of 100pc at the distance of $\omega$ Centauri.
(b): IRAS 100$\micron$ chart overlaid with Leon et al. (2000) tidal tail contours, darker shading indicates a
greater amount of dust along the line of sight.  The total range of color-excess is of order 
$\Delta E(B-V) = 0.18$ mag.
(c): Linear contoured surface density plot of $\omega$ Centauri showing apparent tidal tails
in the non-dereddened 2MASS data.  Dashed lines denote all contours beneath the mean background level.
(d): Surface density plot of $\omega$ Cen after differential dereddening; all stages of the analysis
are identical to those used to create panel (c).  Panels a and b are reproduced from Leon,
Meylan and Combes (2000) courtesy of S. Leon and European Southern Observatory.\label{fig1}}
\end{figure}

\begin{figure}
\epsscale{0.5}
\plotone{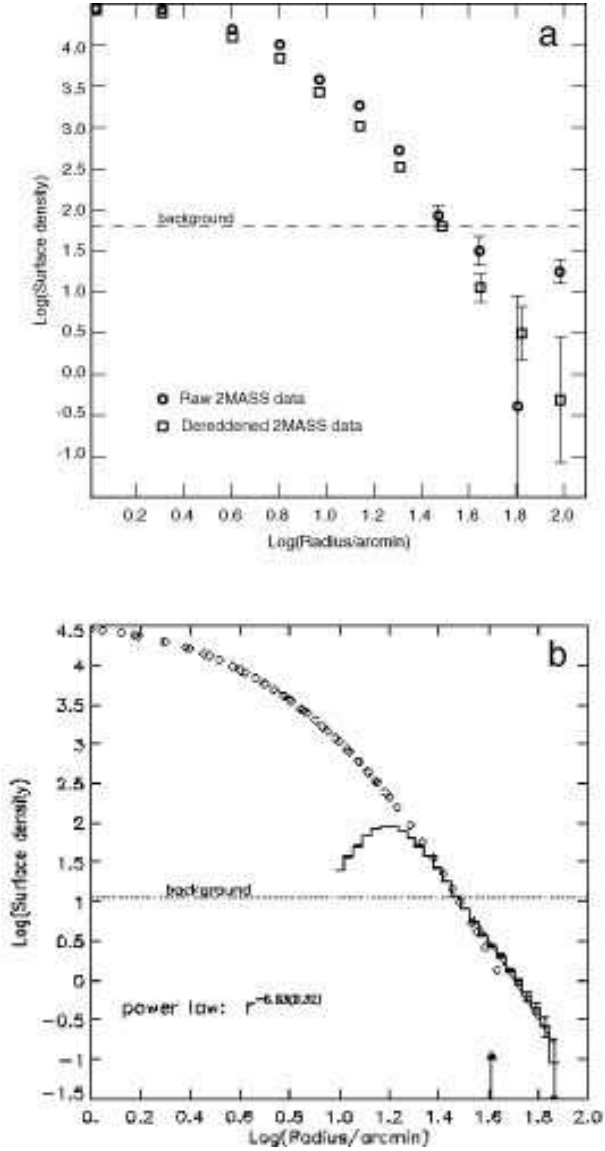}
\caption{(a): Radial surface density profile of $\omega$ Centauri, error bars are derived from Poisson statistics.
Data are shifted vertically to match the
LMC00 profile.  (b): Radial profile reproduced from Leon, Meylan and Combes (2000), Figure
courtesy of S. Leon and European Southern Observatory.  A power law is fit to the data in the
external parts, while the inner surface density profile comes from the data by Trager et al.
(1995).  The vertical arrow indicates the cluster tidal radius.\label{fig2}}
\end{figure}

\begin{figure}
\plotone{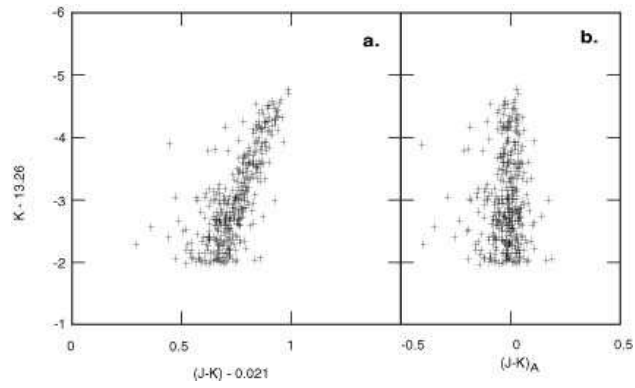}
\caption{RGB of 47 Tuc: before (a) and after (b) removing the mean color magnitude relation.\label{fig3}}
\end{figure}

\begin{figure}
\epsscale{0.8}
\plotone{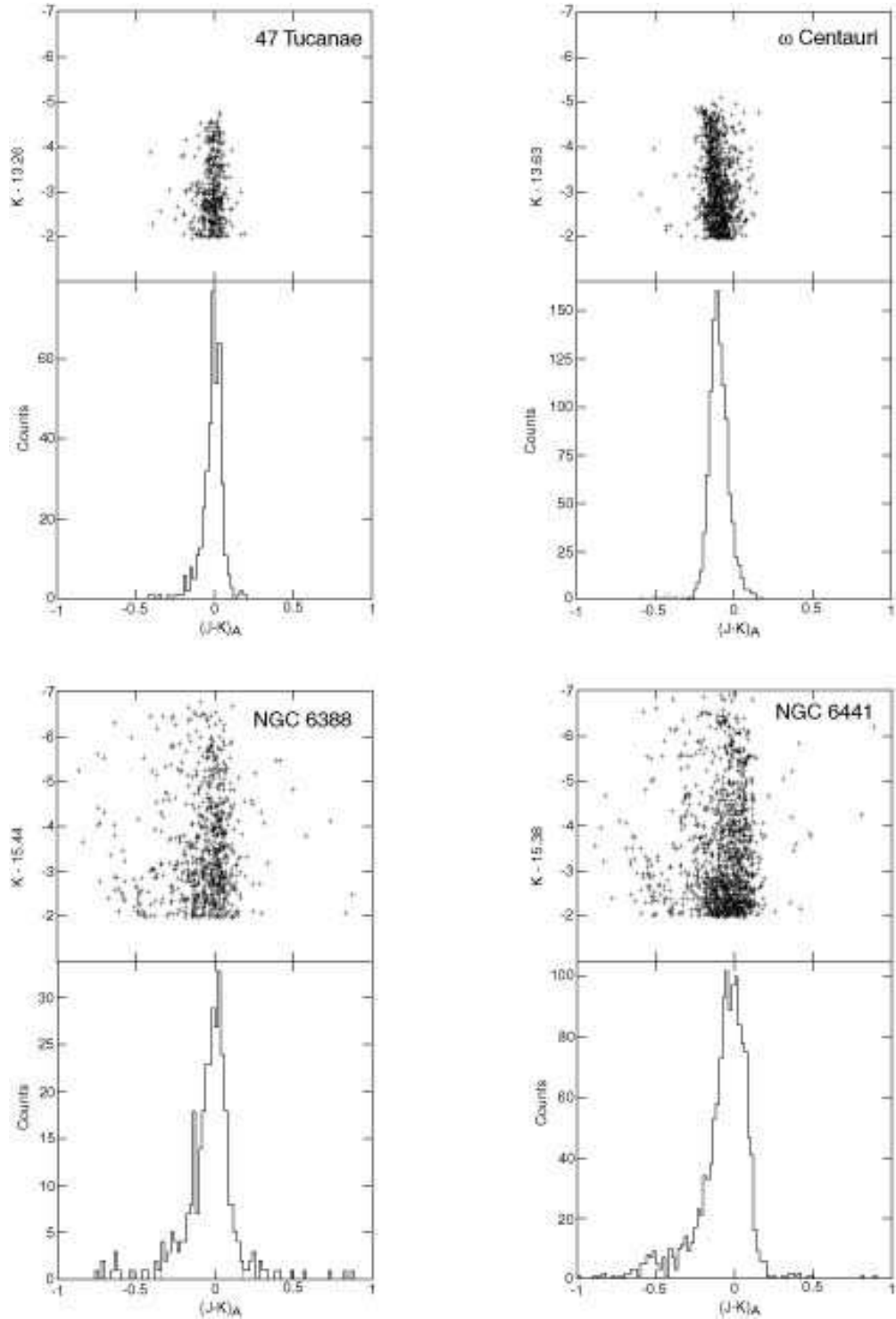}
\caption{Histograms of the RGB for all four clusters.  The field of view is 300 arcseconds.
Note the width of NGC 6388 and NGC 6441.  The $\omega$ Cen RGB is canted to the left because
the color-magnitude relation used to straighten the RGBs was derived from the much more metal
rich 47 Tuc.  Histograms in $(J-H)_A$ are similar in character to the $(J-K)_A$ distributions
shown here.\label{fig4}}
\end{figure}

\begin{figure}
\plotone{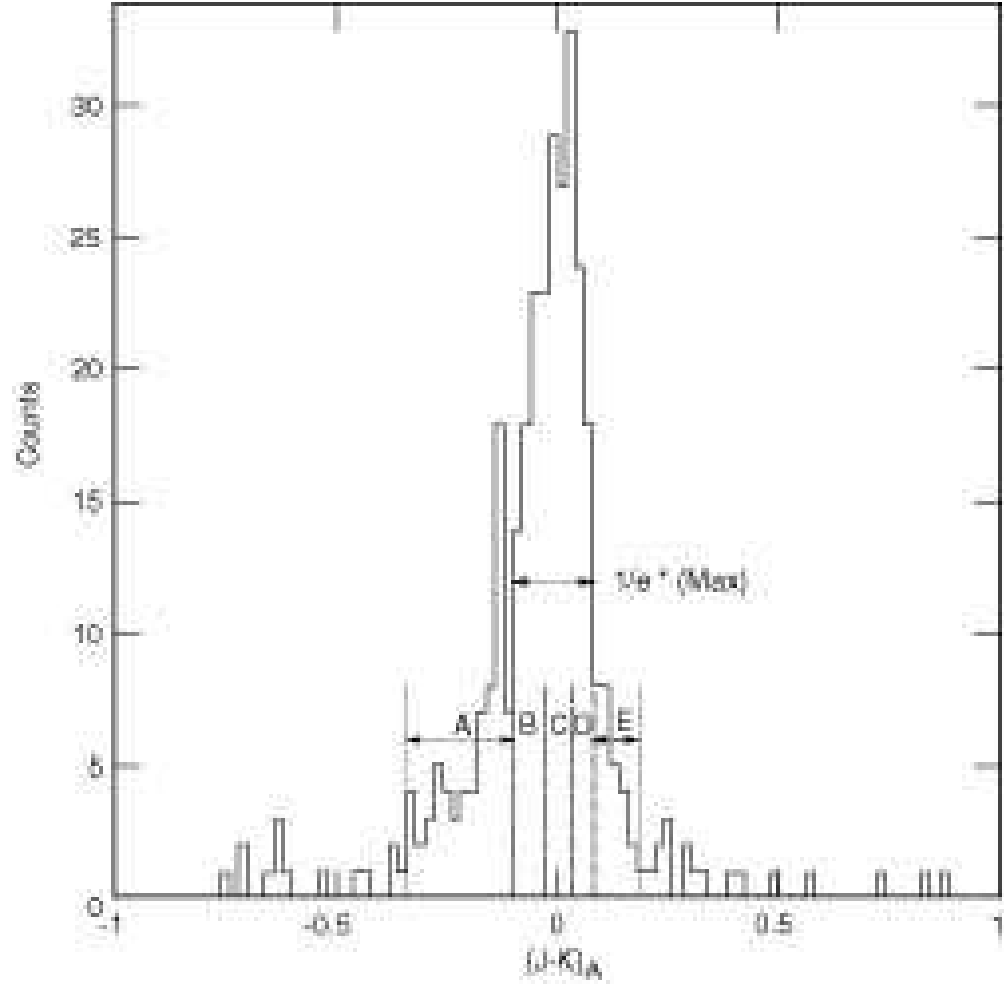}
\caption{Histogram of the RGB of NGC 6388.  The selection criteria for the color bins illustrated
are discussed in Section 4.1.1.\label{fig5}}
\end{figure}

\begin{figure}
\plotone{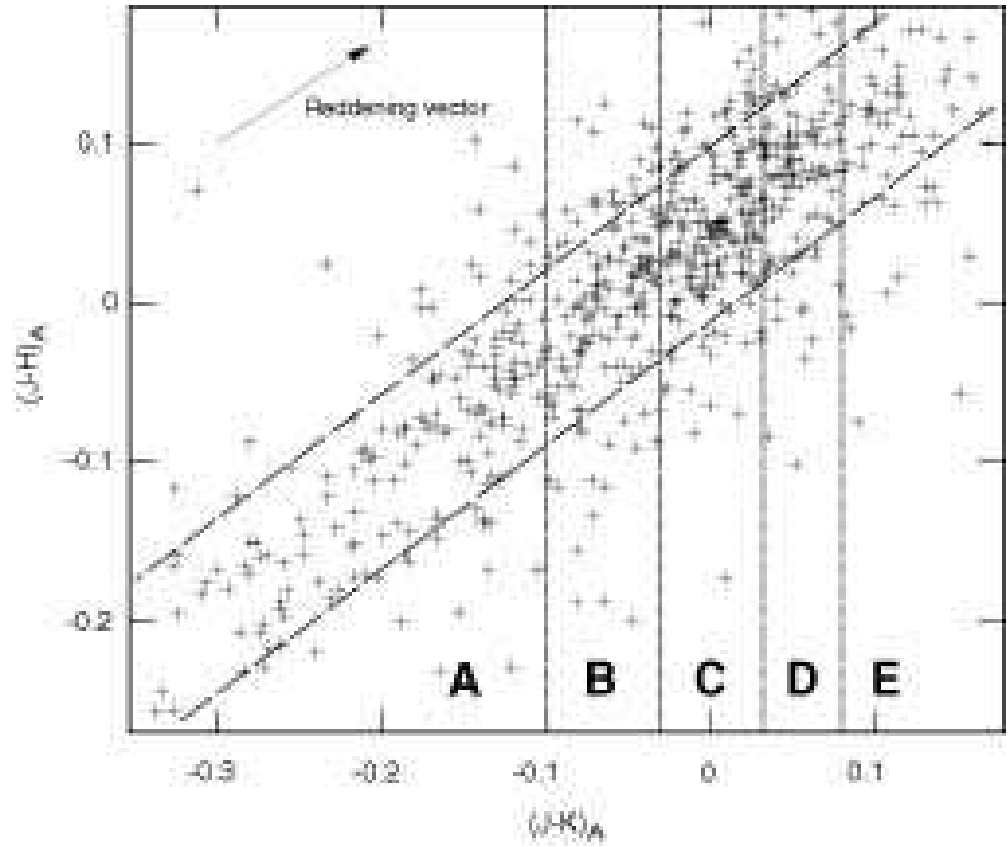}
\caption{Color-color diagram of the RGB of NGC 6388.  Short dashed lines denote boundaries
between $(J-K)_A$ color bins, and long dashed lines denote the location of $(J-K)$, $(J-H)$
color-color cuts, as described in Section 4.1.1.  The fiducial reddening vector is given for
comparison.\label{fig6}}
\end{figure}

\begin{figure}
\plotone{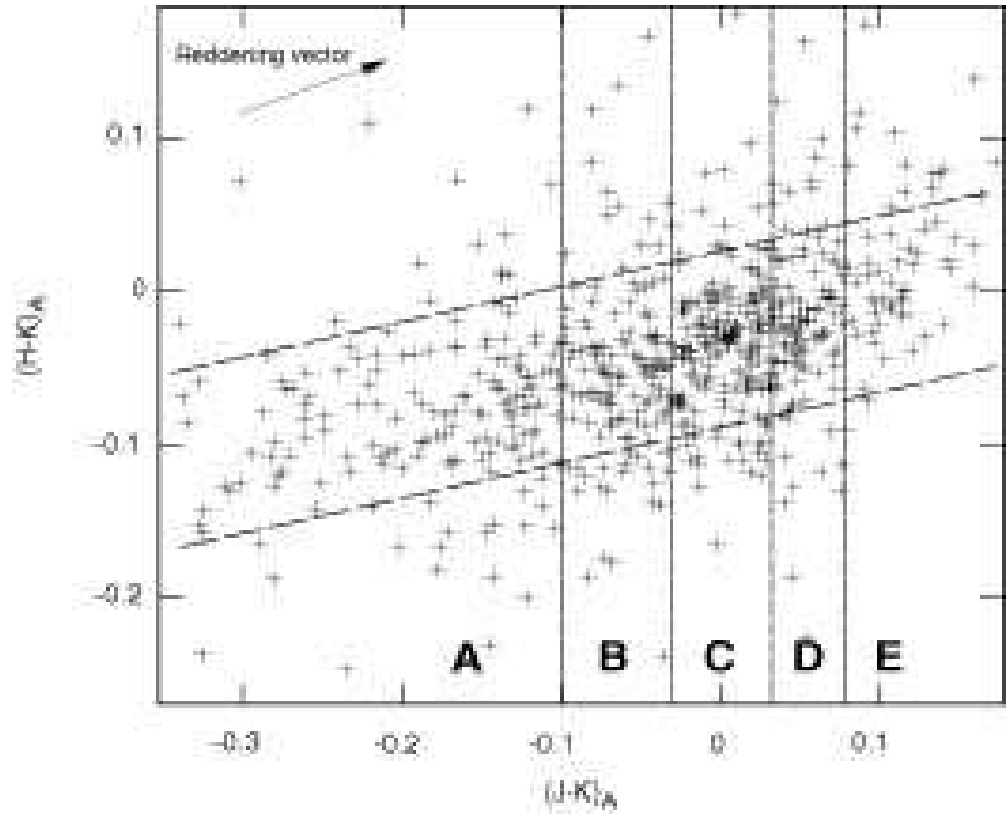}
\caption{Color-color diagram of the RGB of NGC 6388.  Short dashed lines denote boundaries
between bins, and long dashed lines denote the location of $(J-K)$, $(H-K)$ color-color cuts
(as described in Section 4.1.2).\label{fig7}}
\end{figure}

\begin{figure}
\epsscale{0.8}
\plotone{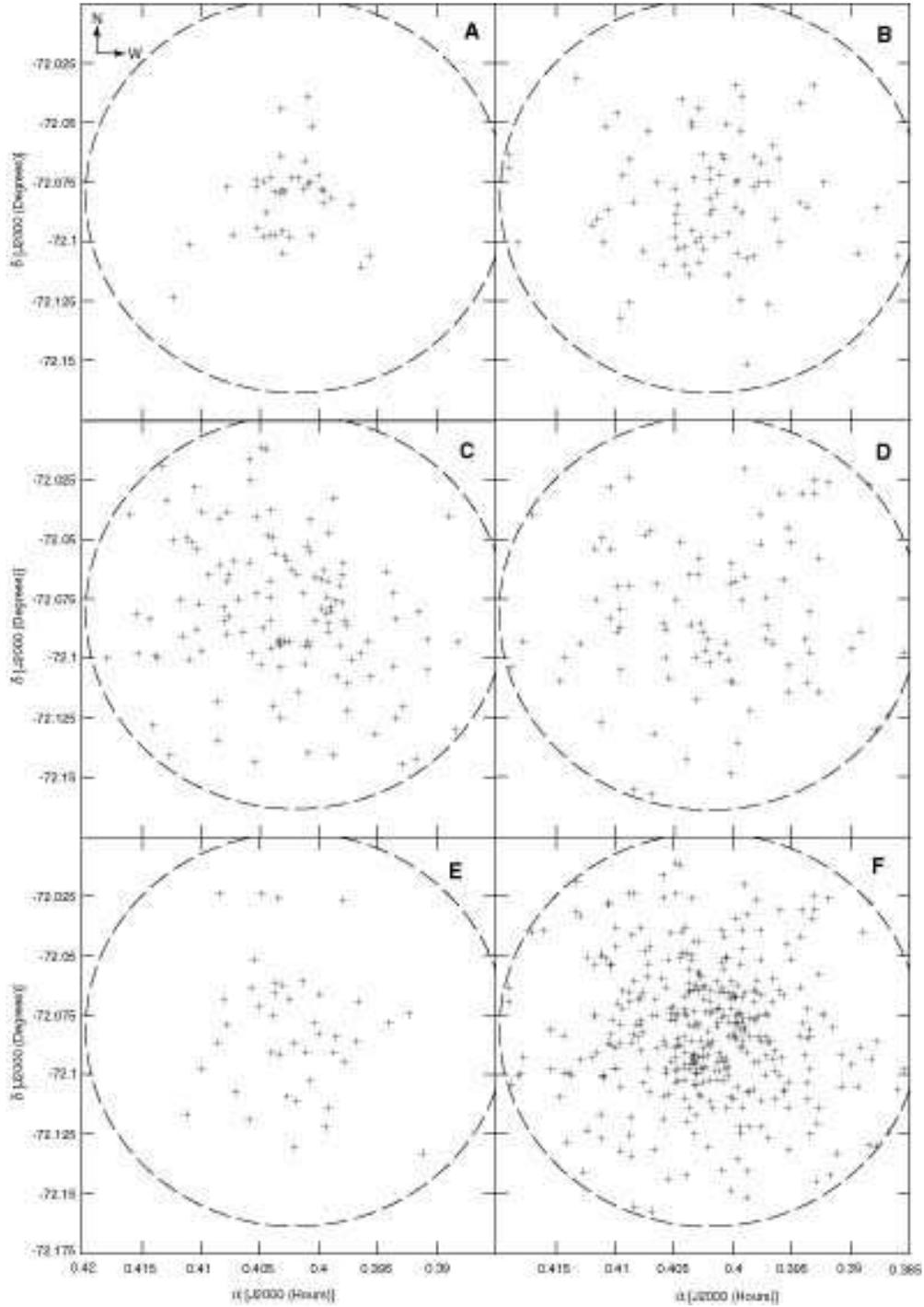}
\caption{Spatial distribution of stars within each bin for the cluster 47 Tucanae.  A: -0.201
$\leq (J-K)_A$ $<$ -0.082, B: -0.082 $\leq (J-K)_A$ $<$ -0.031, C: -0.031 $\leq (J-K)_A$ $<$ 0.011,
D: 0.011 $\leq (J-K)_A$ $<$ 0.042, E: 0.042 $\leq (J-K)_A$ $<$ 0.098, F: All previous bins.
The $(J-H)_A$ sky distribution is similar.  The dashed circle denotes our field of view.\label{fig8}}
\end{figure}

\begin{figure}
\epsscale{0.8}
\plotone{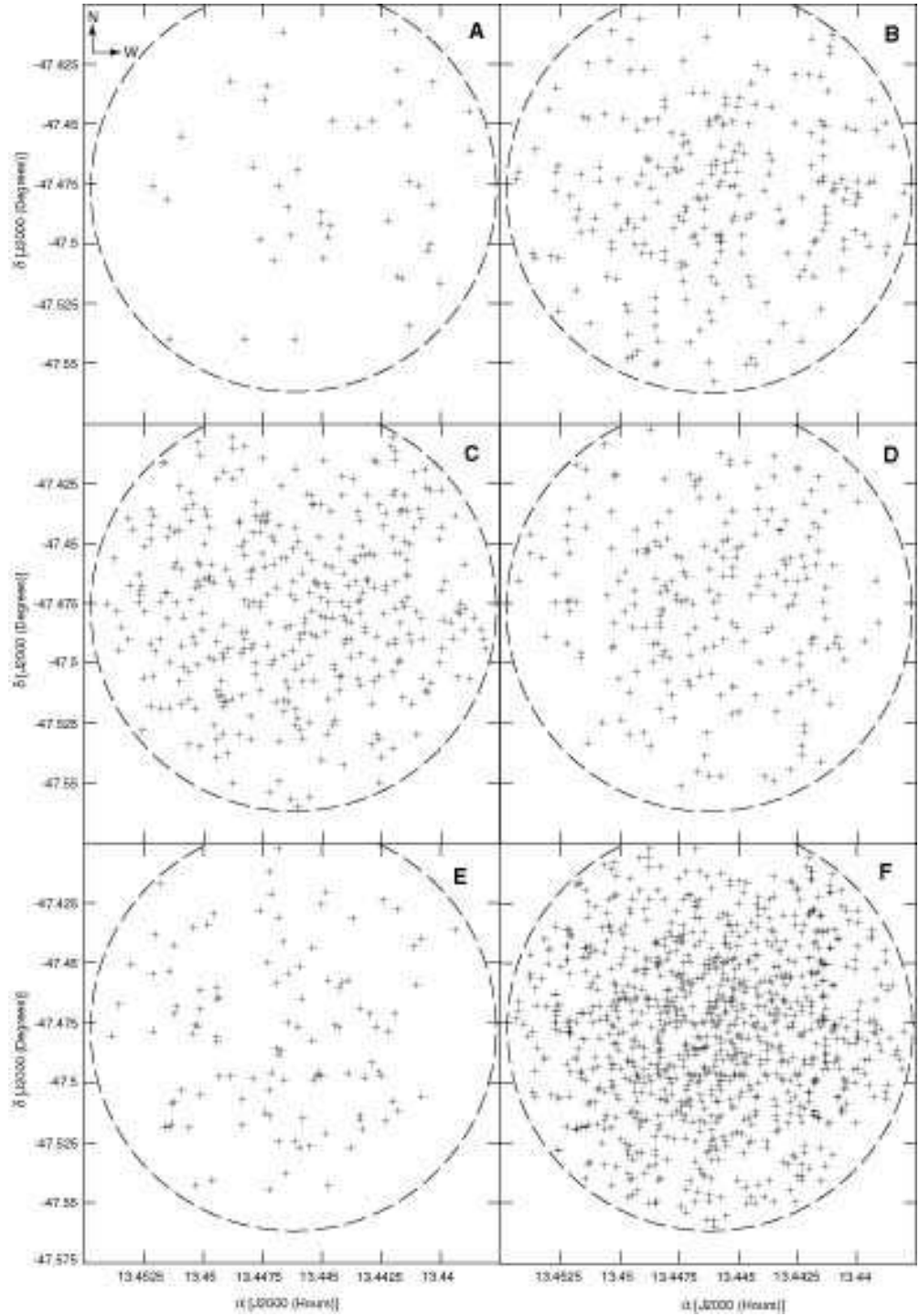}
\caption{Spatial distribution of stars within each bin for the cluster $\omega$ Centauri.
A: -0.22 $\leq (J-K)_A$ $<$ -0.183, B: -0.183 $\leq (J-K)_A$ $<$ -0.129,
C: -0.129 $\leq (J-K)_A$ $<$ -0.075, D: -0.075 $\leq (J-K)_A$ $<$ -0.021,
E: -0.021 $\leq (J-K)_A$ $<$ 0.138, F: All previous bins.\label{fig9}}
\end{figure}

\begin{figure}
\epsscale{0.8}
\plotone{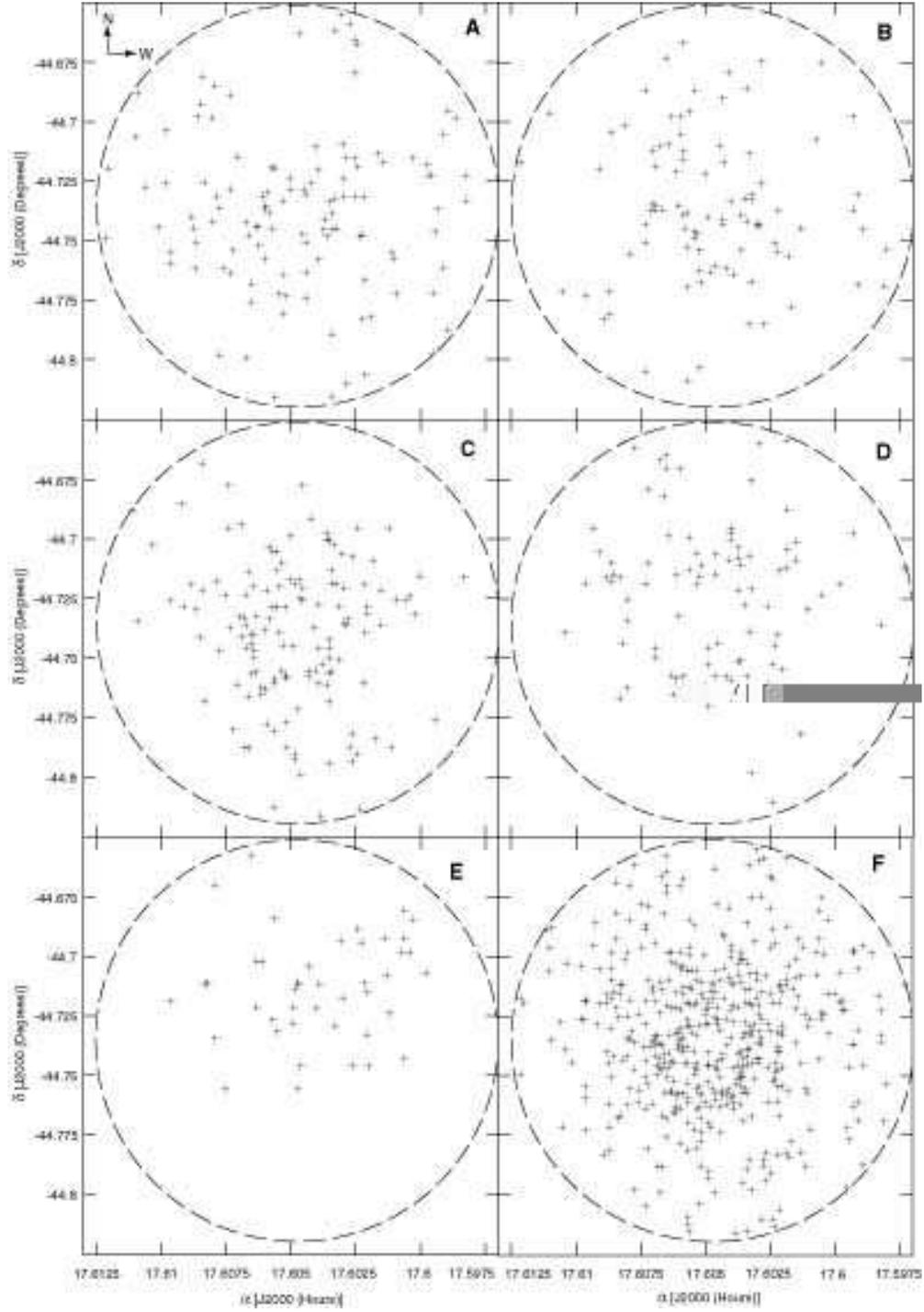}
\caption{Spatial distribution of stars within each bin for the cluster NGC 6388.
A: -0.341 $\leq (J-K)_A$ $<$ -0.101, B: -0.101 $\leq (J-K)_A$ $<$ -0.03,
C: -0.03 $\leq (J-K)_A$ $<$ 0.03, D: 0.03 $\leq (J-K)_A$ $<$ 0.08,
E: 0.08 $\leq (J-K)_A$ $<$ 0.179, F: All previous bins.\label{fig10}}
\end{figure}

\begin{figure}
\epsscale{0.8}
\plotone{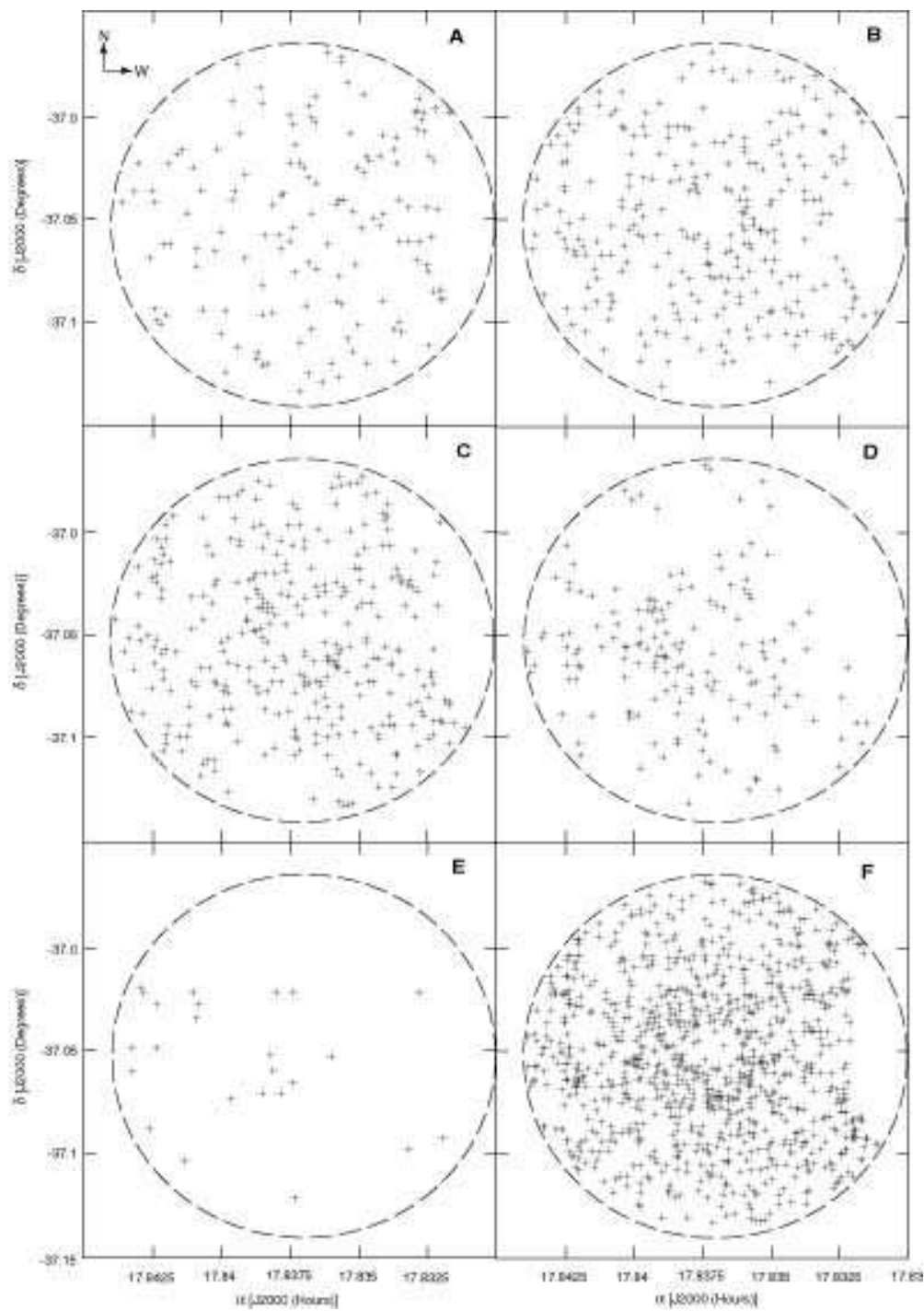}
\caption{Spatial distribution of stars within each bin for the cluster NGC 6441.
A: -0.422 $\leq (J-K)_A$ $<$ -0.16, B: -0.16 $\leq (J-K)_A$ $<$ -0.047,
C: -0.047 $\leq (J-K)_A$ $<$ 0.047, D: 0.047 $\leq (J-K)_A$ $<$ 0.12,
E: 0.12 $\leq (J-K)_A$ $<$ 0.198, F: All previous bins.  $(J-H)_A$ distribution is similar.
Stars are absent from the 2MASS catalog in a small region at the west of the field of view
due to contamination by the bright star HD 161892.\label{fig11}}
\end{figure}

\begin{figure}
\epsscale{0.6}
\plotone{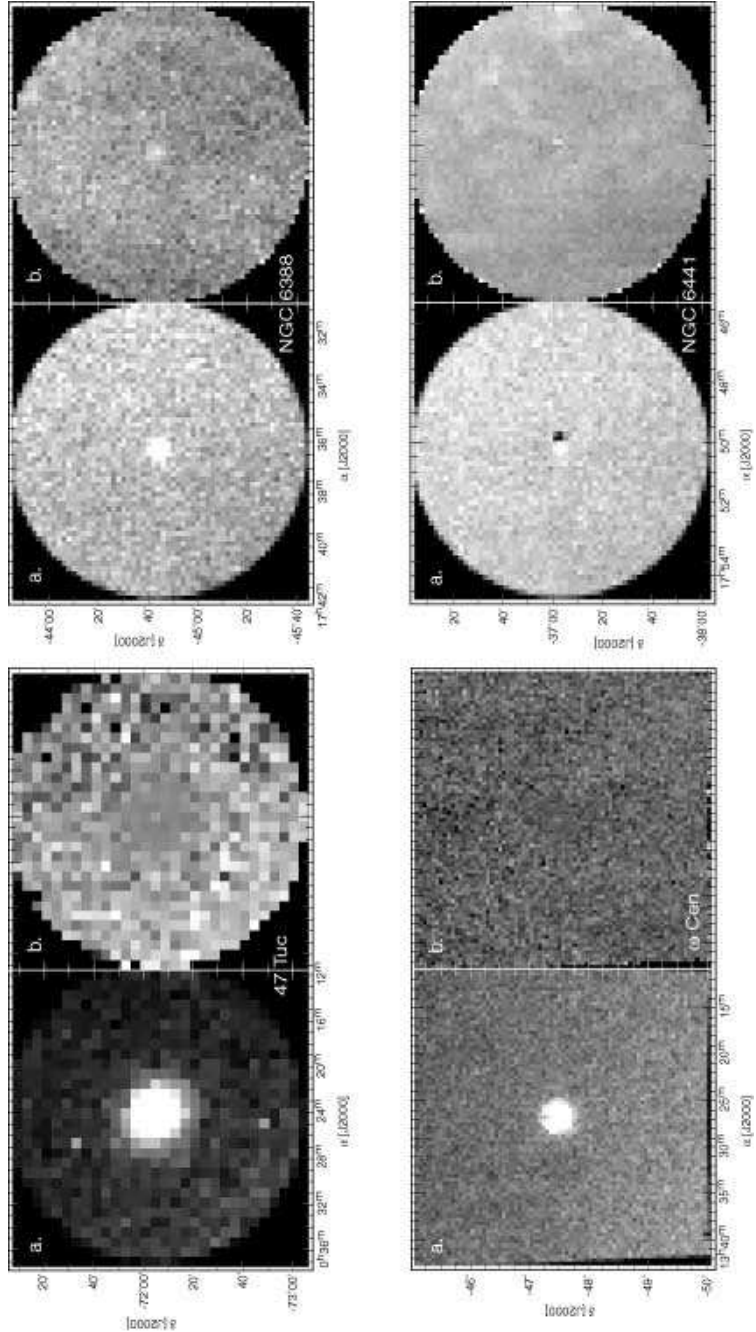}
\caption{2MASS maps of all stars (i.e. both cluster and field stars) within a one degree radius of
each cluster.  In each case, panel
(a) gives the starcount density, and panel (b) the mean stellar color.  Higher starcounts and
redder colors are whiter (see Table 5 for scales).  Vertical stripes in panels (a) are due to
nightly variations in the sensitivity of the 2MASS survey.  Note that the fields of view in
this figure are on the order of square degrees, while in Figures 8-11 the fields of view are
on the order of square arcminutes.  The larger pixel size for 47 Tuc was required to obtain
a reasonable signal to noise ratio, due to low starcounts in this region of high Galactic
latitude.\label{fig12}}
\end{figure}

\begin{figure}
\epsscale{0.9}
\plotone{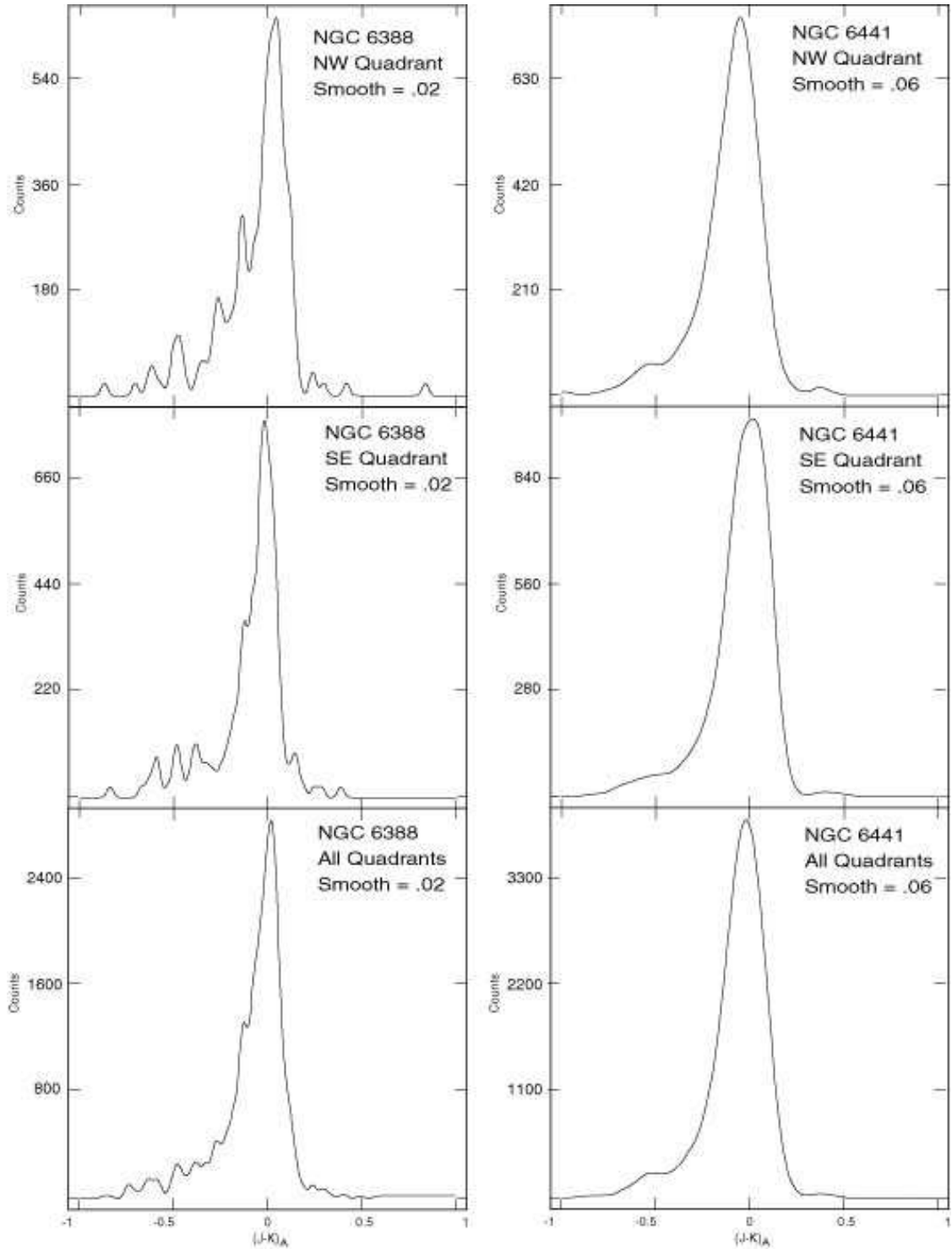}
\caption{$(J-K)_A$ distributions for different regions of NGC 6388 and 6441, smoothed by a
normalized Gaussian kernel of half-maximum width 0.02 and 0.06 respectively.  The vertical scale
on each panel is defined such that the integrated area under the curve is equal to the number of
stars in the given quadrant.  Note the
offset between histograms in NW and SE quadrants for both clusters.\label{fig13}}
\end{figure}

\begin{figure}
\plotone{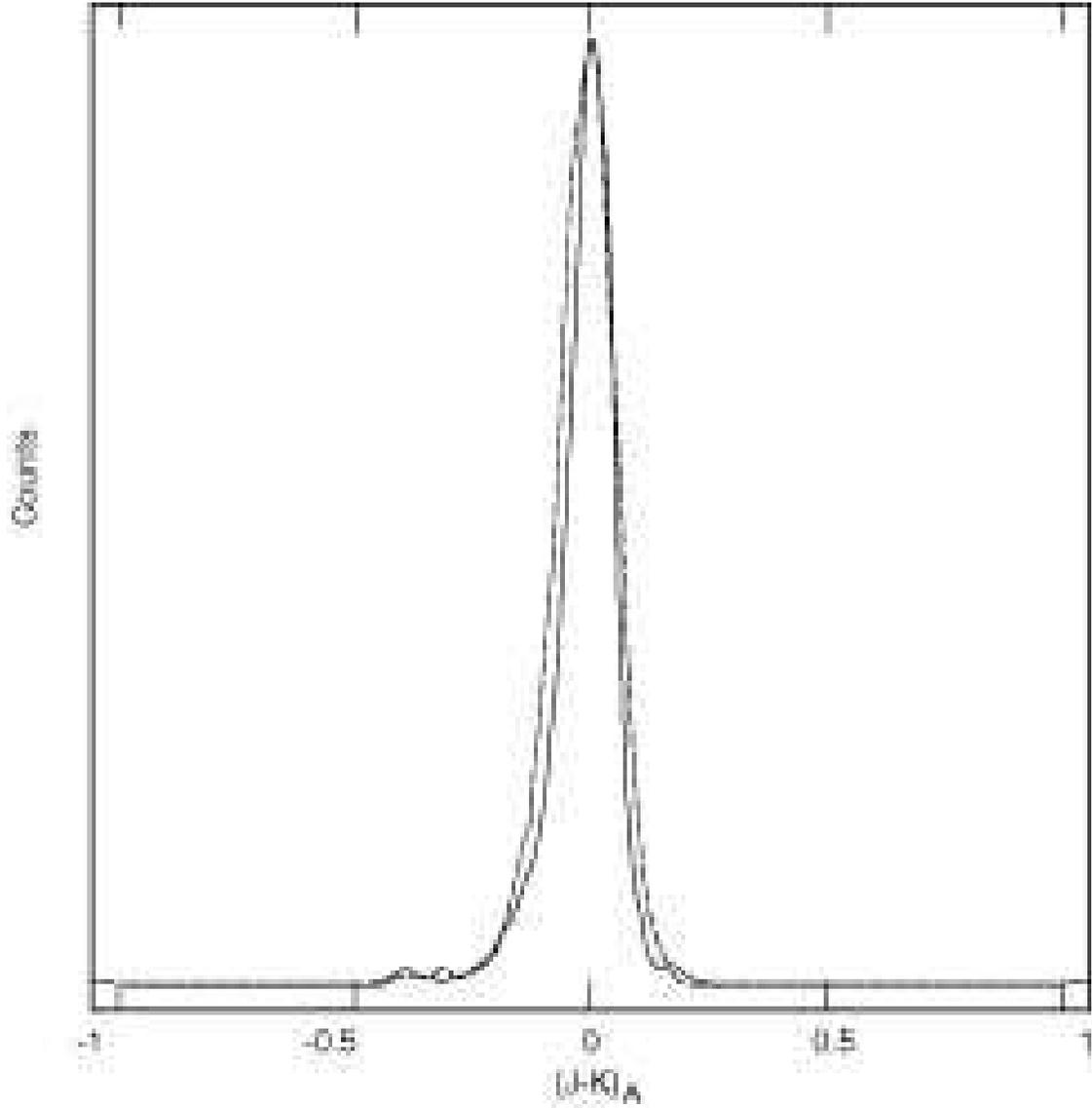}
\caption{$(J-K)_A$ Distributions for 47 Tucanae (solid line) and for 47 Tucanae with
$\Delta(J-K)_A=0.061$ magnitudes of artificial differential reddening (dashed line).
The Gaussian smoothing kernel for both curves has a half-maximum width of 0.02.\label{fig14}}
\end{figure}

\clearpage
\begin{deluxetable}{cccc}
\tablewidth{30pc}
\tablecaption{Cluster Parameters from Harris (1996)}
\tablehead{
\colhead{Cluster} & \colhead{[Fe/H]} & \colhead{Core Radius (arcmin)} & \colhead{Tidal Radius (arcmin)}}
\startdata
47 Tucanae & -0.76 & 0.44 & 47.25\nl
$\omega$ Centauri & -1.62 & 2.58 & 44.85\nl
NGC 6388 & -0.60 & 0.12 & 6.21\nl
NGC 6441 & -0.53 & 0.11 & 8.00\nl
\enddata
\end{deluxetable}

\begin{deluxetable}{cccccccc}
\tablewidth{44pc}
\tablecaption{Extinction Parameters}
\tablecomments{$A_V$, $(m-M)_V$, \& $E(B-V)$ from Harris (1996), all other parameters calculated
according to Cardelli, Clayton, and Mathis (1989).  Note that differential dereddening in the field of
$\omega$ Cen is performed according to the Schlegel et al. (1998) reddening maps, which give an average
color-excess of $E(B-V) = 0.13$ in a $4^{\circ} \times 4^{\circ}$ field around the cluster.}
\tablehead{
\colhead{Cluster} & \colhead{$A_V$} & \colhead{$(m-M)_V$}
& \colhead{$(m-M)_K$}  & \colhead{$E(B-V)$} & \colhead{$E(J-K)$}
& \colhead{$E(J-H)$} & \colhead{$E(H-K)$}}
\startdata
47 Tucanae & 0.124 & 13.37 & 13.26 & 0.04 & 0.021 & 0.011 & 0.009\nl
$\omega$ Centauri & 0.299 & 13.92 & 13.63 & 0.12 & 0.060 & 0.033 & 0.027\nl
NGC 6388 & 1.240 & 16.54 & 15.44 & 0.40 & 0.209 & 0.109 & 0.090\nl
NGC 6441 & 1.395 & 16.62 & 15.38 & 0.45 & 0.234 & 0.123 & 0.102\nl
\enddata
\end{deluxetable}

\begin{deluxetable}{cccc}
\tablewidth{18pc}
\tablecaption{RGB Straightening Coefficients}
\tablecomments{$a$, $b$, \& $c$ are the coefficients of the zeroeth, first, and second order terms
respectively.}
\tablehead{
\colhead{Color} & \colhead{$a$} & \colhead{$b$} & \colhead{$c$}}
\startdata
$(J-K)$ & 0.5436 & -0.0361 & 0.0105\nl
$(J-H)$ & 0.5388 & 0.0066 & 0.0120\nl
$(H-K)$ & 0.0513 & -0.0153 & 0.0021\nl
\enddata
\end{deluxetable}

\begin{deluxetable}{ccc}
\tablewidth{14pc}
\tablecaption{Color-color Elimination Parameters}
\tablecomments{$a$ \& $b$ are the coefficients of the zeroeth and first order polynomial terms
respectively.  $(J-K)_A$ is the abscissa, $(J-H)_A$ the ordinate.}
\tablehead{
\colhead{Cluster} & \colhead{$a$} & \colhead{$b$}}
\startdata
47 Tuc & 0.012 & 0.779\nl
$\omega$ Cen & -0.019 & 0.632\nl
NGC 6388 & 0.044 & 0.766\nl
NGC 6441 & 0.047 & 0.755\nl
\enddata
\end{deluxetable}

\begin{deluxetable}{ccccc}
\tablewidth{32pc}
\tablecaption{Starcount and Reddening Parameters}
\tablecomments{Parameters define the scales used in Fig. 12.}
\tablehead{
\colhead{Cluster} & \colhead{Min Counts} & \colhead{Max Counts} & \colhead{Min $(J-K)$} &
\colhead{Max $(J-K)$}}
\startdata
47 Tuc & --- & --- & --- & ---\nl
$\omega$ Cen & 0 & 5 & 0.4 & 1.0\nl
NGC 6388 & 0 & 15 & 0.4 & 1.0\nl
NGC 6441 & 0 & 30 & 0.4 & 1.2\nl
\enddata
\end{deluxetable}

\begin{deluxetable}{ccccc}
\tablewidth{20pc}
\tablecaption{KS Statistics- 47 Tuc}
\tablecomments{KS statistics shown for correlation significance between the two given quadrants
(probability that the two samples are drawn from the same parent distribution).}
\tablehead{
\colhead{} & \colhead{NE} & \colhead{NW} & \colhead{SE} & \colhead{SW}}
\startdata
NE & --- & 0.599 & 0.228 & 0.256 \nl
NW & 0.599 & --- & 0.539 & 0.116 \nl
SE & 0.228 & 0.539 & --- & 0.629 \nl
SW & 0.256 & 0.116 & 0.629 & --- \nl
\enddata
\end{deluxetable}

\begin{deluxetable}{ccccc}
\tablewidth{20pc}
\tablecaption{KS Statistics- $\omega$ Cen}
\tablecomments{See Table 6 for description.}
\tablehead{
\colhead{} & \colhead{NE} & \colhead{NW} & \colhead{SE} & \colhead{SW}}
\startdata
NE & --- & 0.583 & 0.034 & 0.590 \nl
NW & 0.583 & --- & 0.234 & 0.997 \nl
SE & 0.034 & 0.234 & --- & 0.260 \nl
SW & 0.590 & 0.997 & 0.260 & --- \nl
\enddata
\end{deluxetable}

\begin{deluxetable}{ccccc}
\tablewidth{20pc}
\tablecaption{KS Statistics- NGC 6388}
\tablecomments{See Table 6 for description.}
\tablehead{
\colhead{} & \colhead{NE} & \colhead{NW} & \colhead{SE} & \colhead{SW}}
\startdata
NE & --- & 0.533 & $3 \times 10^{-6}$ & 0.005 \nl
NW & 0.533 & --- & 0.0001 & 0.006 \nl
SE & $3 \times 10^{-6}$ & 0.0001 & --- & 0.471 \nl
SW & 0.005 & 0.006 & 0.471 & --- \nl
\enddata
\end{deluxetable}

\begin{deluxetable}{ccccc}
\tablewidth{20pc}
\tablecaption{KS Statistics- NGC 6441}
\tablecomments{See Table 6 for description.}
\tablehead{
\colhead{} & \colhead{NE} & \colhead{NW} & \colhead{SE} & \colhead{SW}}
\startdata
NE & --- & $4 \times 10^{-6}$ & 0.143 & 0.097 \nl
NW & $4 \times 10^{-6}$ & --- & $9 \times 10^{-7}$ & 0.001 \nl
SE & 0.143 & $9 \times 10^{-7}$ & --- & 0.001 \nl
SW & 0.097 & 0.001 & 0.001 & --- \nl
\enddata
\end{deluxetable}

\end{document}